\documentstyle[11pt]{article}
\begin{document}

\title{Algebraic approach to the Hulthen potential}
\author{Mohammad R. Setare,\thanks{E-mail:rezakord@ipm.ir} \\ Ebrahim
Karimi,\thanks{E-mail:e\_karimi@uok.ac.ir}\\
{Department of Science, University of Kurdistan, Sanandaj, Iran}}
\maketitle
\begin{abstract}
In this paper the energy eigenvalues and the corresponding
eigenfunctions are calculated for Hulthen potential. Then we obtain
the ladder operators and show that these operators satisfy $SU(2)$
commutation relation.
\end{abstract}
\newpage

\section{Introduction}
In the recent years, Lie algebraic methods have been the subject of
the interest in many of fields of physics. For example the algebraic
methods provide a way to obtain wave functions of polyatomic
molecules~\cite{{I1},{I2},{I3},{I4},{I5},{I6}}. These methods
provide a description to Dunham-type expansions and to force-field
variational methods~\cite{I7}. It is clear that systems displaying a
dynamical symmetry can be treated by algebraic
methods~\cite{{D14},{D15},{D16},{D17}}. To see the ladder operators
of a quantum system with some important potentials such as Morse
potential the P\"{o}schel-Teller one, the pseudo harmonic one, the
infinitely square-well one and other quantum systems refer
to~\cite{{D1},{D2},{D3},{D4},{D5},{D6},{D7},{D8},{D9},{D10},{D11}}.\\
We know that the symmetry and degeneracy of the states of a system
are associated with each other. For example, a system that possesses
rotational symmetry is usually degenerate with respect to the
direction of the angular momentum, i.e. with respect to the
eigenvalues of a particular component. Beyond the degeneracies
arising, say, in rotational symmetry there is the possibility of
degeneracies of different origin. Such degeneracies are to be
expected whenever the Shrodinger equation can be solved in more than
one way, either in different coordinate system, or in a single
coordinate system which can be oriented in different directions.
From our present considerations we should expect these degeneracies
to be associated with some symmetry, too. The nature of these
symmetries is not geometrical. They are called dynamical symmetries,
since they are the consequence of particular forms of the Shrodinger
equation or of the classical force law.
\\
In this paper we study the dynamical symmetry for the Hulthen
potential, by another algebraic approach. The Hulthen potential
\cite{h1,h2} is one of the important short-range potentials in
physics.  This potential is a special case of the Eckart potential
\cite{eckart} which has been widely used in several branches of
physics and its bound-state and scattering properties have been
investigated by a variety of techniques (see e.g., \cite{varshni}
and references therein).
 We establish the creation and \mbox{annihilation}
operators directly from the eigenfunctions for this system, and that
the ladders operators construct the dynamical algebra $SU(2)$.

\section{Schr\"{o}dinger equation with Hulthen Potential}
The Hulthen potential has the following form~\cite{h1,h2,F}
\begin{equation}\label{1}
    V(r)=-V_{0}\frac{e^{-\frac{r}{a}}}{1-e^{-\frac{r}{a}}}
\end{equation}
where $V_{0}=Ze^2$ and $a$ are constant parameters. If the potential
is used for atoms, the $Z$ is identified with the atomic number. In
order to calculate the energy eigenvalues and the corresponding
eigenvalues and the corresponding eigenfunctions, the potential
function given by Eq.(\ref{1}) is substituted into the
Schr\"{o}dinger equation:
\begin{equation}\label{2}
    (-\frac{\hbar^2}{2M}\frac{d^2}{dr^2}-V_{0}\frac{e^{-\frac{r}{a}}}{1-e^{-\frac{r}{a}}})
    \psi_{n}(r)=E_{n}\psi_{n}(r)
\end{equation}
By change of coordinate as $x=\frac{r}{a}$, and introducing the
following parameters
\begin{equation}\label{3}
    \varepsilon_{m}=\frac{E_{m}}{V_{0}},
    ~~~~~~~~~~\frac{\hbar^2}{2M}a^2=V_{0}
\end{equation}
Now, we can rewrite Eq.(\ref{2}) as
\begin{equation}\label{4}
    (\frac{d^2}{dx^2}+\frac{e^{-x}}{1-e^{-x}})
    \psi_{n}(x)=-\varepsilon_{n}\psi_{n}(x)
\end{equation}
We would like to consider the bound states with
\begin{equation}\label{5}
    \varepsilon_{n}=-s^2
\end{equation}
We rewrite Eq.(\ref{4}) by using a new variable of the form
$y=e^{-x}$,
\begin{equation}\label{6}
    \{y^2\frac{d^2}{dy^2}+y\frac{d}{dy}+(-s^2+\frac{y}{1-y})\}\psi_{n}(y)=0
\end{equation}
The boundary  conditions areas following
\begin{equation}\label{7}
    \phi_{n}(y)|_{y=0}=0
\end{equation}
\begin{equation}\label{8}
    \phi_{n}(y)|_{y=1}=0
\end{equation}

The solution of Eq.(\ref{6}) are as follow:
\begin{eqnarray}\label{9}
    \phi_{n}(y)&=& y^s(1-y)\{A\, _{2}F_{1}(s+1+\sqrt{s^2+1},s+1-\sqrt{s^2+1},2s+1;y)\cr
    &+& B y^{-2s}\, _{2}F_{1}(-s+1+\sqrt{s^2+1},-s+1-\sqrt{s^2+1},-2s+1;y) \}
\end{eqnarray}
where $A$, and $B$ are constant coefficients. Using the first
boundary condition~(\ref{7}), $B=0$. Now to consider the second
boundary condition~(\ref{8}), we expand the hypergeometric function
$_{2}F_{1}(a,b,c;y)$ near the $y=1$~\cite{L}
\begin{eqnarray}\label{10}
    &&_{2}F_{1}(s+1+\sqrt{s^2+1},s+1-\sqrt{s^2+1},2s'+1;y)=\frac{\Gamma(2s'+1)\Gamma(\epsilon-1)}{\Gamma(s-\sqrt{s^2+1}+\epsilon)\Gamma(s+\sqrt{s^2+1}+\epsilon)}
    \cr\cr
    &&\,{}_{2}F_{1}(s+1+\sqrt{s^2+1},s+1-\sqrt{s^2+1},2s'+1;1-y)\cr\cr
    &&+(1-y)^{\epsilon-1}\frac{\Gamma(2s'+1)\Gamma(\epsilon-1)}{\Gamma(s+1+\sqrt{s^2+1})\Gamma(s+1-\sqrt{s^2+1})}\cr\cr
    &&\times\,{}_{2}F_{1}(s-\sqrt{s^2+1}+\epsilon,s+\sqrt{s^2+1}+\epsilon,\epsilon;1-y)
\end{eqnarray}
where $\epsilon=2(s'-s)$. Also we have
\begin{equation}\label{11}
    {}_{2}F_{1}(a,b,\epsilon;1-y)=1+\frac{ab}{\epsilon}\,(1-y)+\cdots
\end{equation}
Now we consider the limit $\epsilon\rightarrow 0$ of Eq.~(\ref{10}),
the first term of (\ref{10}) in this limit will be finite, if
\begin{equation}\label{12}
    s-\sqrt{s^2+1}=-n,~~~~~~~~~~n=0,1,2,\cdots
\end{equation}
then we have following relation for the first term of (\ref{10}) in
the limit $\epsilon\rightarrow 0$
\begin{equation}\label{13}
    \lim_{\epsilon\rightarrow
    0}\frac{\Gamma(\epsilon-1)}{\Gamma(\epsilon-n)}=(-1)^{n+1}n!
\end{equation}
Due to the relation (\ref{12}), in second term of Eq.~(\ref{10}),
$\Gamma(s+1-\sqrt{s^2+1})=(-n)!$  whit is infinite unless
$s-\sqrt{s^2+1}=0$, in this case using Eq.~(\ref{11}), we can write
the second term of~(\ref{10}):
\begin{equation}\label{14}
    \lim_{{\epsilon\rightarrow 0},\,{y\rightarrow 1}}(1-y)^{-1}
    \frac{\Gamma(2s+1)}{\Gamma(s+1+\sqrt{s^2+1})}\,\{ 1+
    \frac{\epsilon(s+1+\sqrt{s^2+1}+\epsilon)}{\epsilon}\,(1-y)+\cdots\}
    \propto(1-y)^{-1}
\end{equation}
Then above relation is divergent when $y\rightarrow 1$, then the $n$
could not take the zero value, in another term
\begin{equation}\label{15}
      s-\sqrt{s^2+1}=-n,~~~~~~~~~~n=1,2,\cdots
\end{equation}
Finally we obtain following expression for the wave function
\begin{equation}\label{16}
      \psi_{n}(y)=N_{n}y^s(1-y){}_{2}F_{1}(2s+1+n,1-n,2s+1;y)
\end{equation}
where the normalization factor is given by
\begin{equation}\label{17}
    N_{n}=\{\int_{0}^{1}dy\, y^{2s}(1-y)^{2}{}_{2}F_{1}^{2}(2s+1+n,1-n,2s+1;y)\}^{-\frac{1}{2}}
\end{equation}
some values of $N_{n}$ for different $n$ are given in the
table~(\ref{tab1}). Using Eqs.~(\ref{5},\ref{15}) we have
\begin{equation}\label{18}
    \varepsilon_{n}=-s^2=-(\frac{n^2-1}{2n})^2
\end{equation}
then using Eq.~(\ref{3}), one can determine the energy eigenvalues
$E_{n}$ as
\begin{equation}\label{19}
   E_{n}=-V_{0}(\frac{n^2-1}{2n})^2,~~~~~~~n=1,2,\cdots
   \end{equation}
   in this case the energy level is not equidistant.
\section{Ladder operators for the Hulthen potential}
In this section we address the problem of finding creation and
annihilation operators for the Hulthen wave function~(\ref{16}),
namely, we intend to find different operators $\hat{L}_{\pm}$ with
following property:
\begin{equation}\label{20}
    \hat{L}_{\pm}\,\psi_{n}(y)=l_{\pm}\,\psi_{n}(y)
\end{equation}
we start by establishing the action of the differential operator
$\frac{d}{dy}$ on the Hulthen wave functions
\begin{eqnarray}\label{21}
    \frac{d}{dy}\psi_{n}(y)&=&\frac{d}{dy}(N_{n}y^s(1-y)\,{}_{2}F_{1}(2s+1+n,1-n,2s+1;y))\cr
    &=&(\frac{s}{y}-\frac{1}{1-y})\psi_{n}(y)+
    N_{n}y^s(1-y)\frac{d}{dy}({}_{2}F_{1}(2s+1+n,1-n,2s+1;y))
\end{eqnarray}
To obtain the wanted result, we use the following relations~\cite{L}
\begin{eqnarray}\label{22}
    &&\frac{d}{dx}\,{}_{2}F_{1}(a,b,c;x)=\frac{a b}{c}\,{}_{2}F_{1}(a+1,b+1,c+1;x)\\
    \label{23}&&\frac{a}{c}\,x\,
    {}_{2}F_{1}(a+1,b+1,c+1;x)={}_{2}F_{1}(a,b+1,c;x)-\,{}_{2}F_{1}(a,b,c;x)\\
    \label{24}&&(a-b)\,{}_{2}F_{1}(a,b,c;x)+a\,{}_{2}F_{1}(a+1,b,c;x)+b\,{}_{2}F_{1}(a,b+1,c;x)=0\\
    \label{25}&&(a-b)(1-x)\,{}_{2}F_{1}(a,b,c;x)+(c-a)\,{}_{2}F_{1}(a-1,b,c;x)-(c-a)\,{}_{2}F_{1}(a,b-1,c;x)=0
\end{eqnarray}
Using Eq.~(\ref{22}), we obtain following relation for last term
of~(\ref{21}):
\begin{equation}\label{26}
    \frac{d}{dy}({}_{2}F_{1}(2s+1+n,1-n,2s+1;y))=
    \frac{(2s+1+n)(1-n)}{(2s+1)}{}_{2}F_{1}(2s+1+(n+1),1-n+1,2s+1+1;y)
\end{equation}
Now we use Eqs.(\ref{23},\ref{24},\ref{25}) to satisfy the above
relation
\begin{eqnarray}\label{27}
    &&\frac{d}{dy}({}_{2}F_{1}(2s+1+n,1-n,2s+1;y))=\cr
    &&\frac{(2s+n+1)}{y}\{\frac{(2s+n)}{(1-y)(2s+2n+1)}
    \,{}_{2}F_{1}(2s+1+(n+1),1-(n+1),2s+1;y)\cr
    &&+(\frac{(n+1)}{(1-y)(2s+2n+1)}-1)\,{}_{2}F_{1}(2s+1+(n),1-(n),2s+1;y)
    \}
\end{eqnarray}
By substituting the above realtion in Eq.~(\ref{21}) we obtain
\begin{eqnarray}\label{28}
    \frac{d}{dy}\psi_{n}(y)&=&(\frac{s}{y}-\frac{1}{1-y}+\frac{2s+n+1}{y}(\frac{(n+1)}{(1-y)(2s+2n+1)}-1))\psi_{n}(y)\cr
    &+&(\frac{(2s+n+1)(2s+n)}{y(1-y)(2s+2n+1)})\frac{N_{n}}{N_{n+1}}\psi_{n+1}(y)
\end{eqnarray}
we can rewrite the above equation in the standard form~(\ref{20})
\begin{eqnarray}\label{29}
    [y(1-y)\frac{d}{dy}+y-s(1-y)&+&(1-y)(2s+n+1)(\frac{(n+1)}{(1-y)(2s+2n+1)}-1)]\cr
    \times(\frac{2s+2n+1}{2s+n+1})\psi_{n}(y)&=&(2s+n)\frac{N_{n}}{N_{n+1}}\psi_{n+1}(y)
\end{eqnarray}
therefor we have following relation for the creation operator
\begin{equation}\label{30}
    \hat{L}_{+}=[y(1-y)\frac{d}{dy}+y-s(1-y)+(1-y)(2s+n+1)(\frac{(n+1)}{(1-y)(2s+2n+1)}-1)]
    \times(\frac{2s+2n+1}{2s+n+1})
\end{equation}
satisfying the equation
\begin{equation}\label{31}
    \hat{L}_{+}\,\psi_{n}(y)=l_{+}\,\psi_{n+1}(y)
\end{equation}
with
\begin{equation}\label{32}
    l_{+}=(2s+n)\frac{N_{n}}{N_{n+1}}
\end{equation}
Similarly one can obtain the annihilation operator as
\begin{equation}\label{33}
    \hat{L}_{-}=[-y(1-y)\frac{d}{dy}-y+s(1-y)+(1-y)(n-1)(1-\frac{(2s+n-1)}{(1-y)(2s+2n-1)})]
    \times(\frac{2s+2n-1}{n-1})
\end{equation}
with the following effect over the wave functions:
\begin{equation}\label{34}
    \hat{L}_{-}\,\psi_{n}(y)=l_{-}\,\psi_{n-1}(y)
\end{equation}
where
\begin{equation}\label{35}
    l_{-}=(n)\frac{N_{n}}{N_{n-1}}
\end{equation}
We now study the algebra associated to the operators $\hat{L}_{+}$
and $\hat{L}_{-}$. Based on results (\ref{31},\ref{34}) we can
calculate the commutator $[\hat{L}_{+},\hat{L}_{-}]$:
\begin{equation}\label{36}
    [\hat{L}_{+},\hat{L}_{-}]\psi_{n}(y)=2(n+s)\psi_{n}(y)
\end{equation}
we can define the operator
\begin{equation}\label{37}
    \hat{L}_{0}=(\hat{n}+s)
\end{equation}
where $\hat{n}$ is the number operator
\begin{equation}\label{38}
    \hat{n}\,\psi_{n}(y)=n\,\psi_{n}(y)
\end{equation}
thus the operator $\hat{L}_{0}$ has the following eigenvalue
\begin{equation}\label{39}
    l_{0}=n+s
\end{equation}
Thus the operators $\hat{L}_{\pm}$, $\hat{L}_{0}$ satisfy the
commutation relations:
\begin{equation}\label{40}
    [\hat{L}_{-},\hat{L}_{+}]=2\hat{L}_{0},
    ~~~~~[\hat{L}_{0},\hat{L}_{-}]=-\hat{L}_{-},
    ~~~~~~[\hat{L}_{0},\hat{L}_{+}]=\hat{L}_{+}
\end{equation}
Which correspond to the $SU(2)$ group for the Hulthen potential.
\section {conclusion}
In this paper, we have calculated the exact bound-state energy
eigenvalues and the corresponding eigenfunctions of the Hulthen
potential. We have  shown that  the energy level is not equidistant
in this case. Then we have obtained the raising and lowering
operators and we have shown that $SU(2)$ is the dynamical group
associated with the bounded region of the spectrum.
\appendix

\newpage
\section*{Tables}

\begin{table}[h]
 \begin{center}
  \begin{tabular}{|c|c|}\hline
   $n$ & $N_{n}$  \\   \hline
  $1$ & $\sqrt{3+11 s+12 s^2+4 s^3}$ \\\hline
  $2$ & $\frac{1+2 s}{2}\sqrt{30+47s+24s^2+4s^3}$ \\\hline
  $3$ & $\frac{1+3s+2s^2}{3}\sqrt{105+107s+36s^2+4 s^3}$ \\\hline
  $4$ & $\frac{3+11s+12s^2+4s^3}{12}\sqrt{252+191s+48s^2+4s^3}$ \\
  \hline
\end{tabular}
\caption{\label{tab1}This table shown four normalization factor of
Eq.~(\ref{16}).}
\end{center}
\end{table}


\begin{thebibliography}{5}
\bibitem{I1} O. S. van Roosmalen, A. E. Dieperink and F. Iachello, Chem. Phys.
Lett. {\bf 85}, 52 (1982).
\bibitem{I2} O. S. van Roosmalen, F. Iachello, R.D. Levine
and A. E. Dieperink, J. Chem. Phys. {\bf 79}, 2515 (1983).
\bibitem{I3} O. S. van Roosmalen, I. Benjamin and  R. D. Levine, J. Chem. Phys. {\bf 81}, 5986 (1984).
\bibitem{I4} F. Iachello and S. Oss, Phys. Rev. Lett. {\bf 66}, 2976 (1991).
\bibitem{I5} For a review of Lie algebraic methods in molecular spectroscopy,
see: F. Iachello and R. D. Levine, Algebraic Theory of Molecules;
Oxford University Press: New York, (1995).
\bibitem{I6} F. Iachello and  M. Ibrahim, J. Phys. Chem. A {\bf 102}, 9427-9432 (1998)
\bibitem{I7} F. Iachello and  S. Oss, J. Chem. Phys. {\bf 104}, 6954 (1996)
\bibitem{D14} A. Arima and F. Iachello, Ann. Phys. {\bf 99}, 253 (1974).
\bibitem{D15} A. Perelomov, Generalized Coherent States and their Applications. New York, Springer (1985).
\bibitem{D16} R. G. Wybourne, Classical Groups for Physicists. New York, Wiley (1974).
\bibitem{D17} I. L. Cooper, J. Phys. A: Math. Gen. {\bf 26}, 1601 (1993).
\bibitem{D1} S. H. Dong, R. Lemus and A. Frank, Int. J. Quan. Chem. {\bf 86} 433 (2002).
\bibitem{D2} S. H. Dong and R. Lemus, Int. J. Quan. Chem. {\bf 86}, 265 (2002).
\bibitem{D3} S. H. Dong, Can. J. Phys. {\bf 80}, 129 (2002).
\bibitem{D4} S. H. Dong,  Z. Phys. Chem. {\bf 216}, 103 (2002).
\bibitem{D5} S. H. Dong, and Z. Q. Ma, 2002. Am. J. Phys. {\bf 70}, 520 (2002).
\bibitem{D6} S. H. Dong, and Z. Q. Ma,  Int. J. Mod. Phys. E {\bf 11}, 155 (2002).
\bibitem{D7} S. Dong, and S. H. Dong, Czech. J. Phys. {\bf 52}, 753 (2002).
\bibitem{D8} S. Dong and S. H. Dong,  Int. J. Mod. Phys. E {\bf 11}, 265 (2002).
\bibitem{D9} S. H. Dong,  Appl. Math. Lett. {\bf 16},199 (2003).
\bibitem{D10} S. H. Dong, Computer and Mathematics with application {\bf 47}, 1037 (2004).
\bibitem{D11} S. H. Dong,  G.H. Sun and Y. Tang, Int. J. Mod. Phys. E {\bf 12}, 809 (2003).
\bibitem{h1} L. Hulthen, Ark. Mat. Astron. Fys.{\bf 28A}, 5,
(1942).
\bibitem{h2} L. Hulthen, Ark. Mat. Astron. Fys.{\bf 29B}, 1, (1942).
\bibitem{eckart} C. Eckart, Phys. Rev. {\bf 35}, 1303, (1930).
\bibitem{varshni} Y. P. Varshni, Phys. Rev. {\bf A 41}, 4682, (1990).
\bibitem{F}  S. Fl\"{u}gge, Practical Quantum Mechanics. 2nd Printe. Springer, PP 175-178 (1999).
\bibitem{L} N. N. Lebedev, Special Functions and Their Applications. Translated by Richard A. Silverman,
Dover Publication, Inc., New York, Chapter 5 (1972).

\end{thebibliography}
\end{document}